\documentclass[preprint,prd,showpacs,preprintnumbers,amsmath,amssymb,floatfix]{revtex4}

\usepackage{graphicx}
\usepackage{dcolumn}
\usepackage{bm}
\usepackage{multirow}

\newcommand{\beq}{\begin{equation}}
\newcommand{\eeq}{\end{equation}}
\newcommand{\bea}{\begin{eqnarray}}
\newcommand{\eea}{\end{eqnarray}}
\newcommand{\cM}{{\cal M} }
\newcommand{\cK}{{\cal K}}
\newcommand{\cO}{{\cal O} }
\newcommand{\cC}{{\cal C} }

\begin{document}

\title{Improved Superlinks for Higher Spin Operators}

\author{Robert W. Johnson}
\email{rob.johnson@gatech.edu}
\affiliation{Georgia Institute of Technology, Atlanta, GA 30332, USA}

\date{June 10, 2007}

\begin{abstract}
Traditional smearing or blocking techniques serve well to increase the overlap of operators onto physical states but allow for links orientated only along lattice axes.  Recent attempts to construct more general propagators have shown promise at resolving the higher spin states but still rely on iterative smearing.  We present a new method of superlink construction which creates smeared links from (sparse) matrix multiplications, allowing for gluonic propagation in arbitrary directions.  As an application and example, we compute the positive-parity, even-spin glueball spectrum up to spin 6 for pure gauge SU(2) at $\beta=6$, L\,=\,16, in D\,=\,2+1 dimensions.
\end{abstract}



\maketitle

\section{\label{sec:intro}Introduction}

Operator construction in lattice gauge theory relies on the use of smeared and/or blocked links to improve the overlap onto physical states.  (See \cite{teper-1999-59} for a review of SU(N) gauge theory in 2+1 dimensions.)  Smearing serves to reduce the ultraviolet fluctuations, allowing for a better projection onto the physically smooth wave-functions of glueball (and other) states, and blocking serves to keep the operator size physical as the lattice spacing decreases.  Unfortunately, traditionally smeared links are only defined along axes directions, allowing construction of at most $2(D-1)$-fold symmetric operators, with the rotation symmetries of the lattice timeslice.  In D\,=\,2+1, these operators couple to states with spin $|J|$ congruent {\it modulo} 4, {\it ie} a trial spin 0 operator may contain contributions from spins 0, 4, 8, {\it etc}, and a trial spin 1 operator may contain contributions from spins 1, 3, 5, {\it etc} \cite{johnson-2002-66b,johnson-1999-73}.  To remedy the situation, links along either arbitrary or diagonal directions have been constructed \cite{johnson-2002-66b,meyer-2003-658}, allowing for operators with higher-order symmetry.  These have been successful at resolving the 0/4 and 1/3 ambiguities \cite{johnson-2002-66a,meyer-2003-668} in the 2+1 dimensional pure gauge theory spectrum.

The ``path constructor method'' of \cite{meyer-2003-658} creates blocked links along the diagonal lattice directions for each blocking level, which are then used to define links between arbitrary sites $i$ and $j$ in a given timeslice.  While effective, care must be taken of the details of the path construction for each unique arrangement of $i$ and $j$.  The ``matrix method'' of \cite{johnson-2002-66b,meyer-2003-658} creates the matrix of Green's functions between any and all sites $i$ and $j$ from the inversion of a matrix formed from the link variables of the timeslice, with a ``hopping parameter'' included to control the effective smearing level.  While a single algorithm constructs all the paths at once, the inversion is painfully slow compared to iterative blocking and care must be taken to suppress unwanted torelon contributions.  Note, however, that the correlation of rotated segment operators shown in \cite{meyer-2003-658} is much smoother for the matrix method, implying a better approximation of rotational invariance.  Furthermore, the matrix method appears to be directly applicable to fermion propagator calculations with some minor modification \cite{meyer-2003-658,Montvay:1994cy,Creutz:1983}.  For these reasons, further investigation of the matrix method is warranted.

In this article, we explore a new method of smeared link construction based upon the matrix method but which alleviates its primary difficulties.  Matrix inversion is replaced with matrix multiplications, and torelons are suppressed efficiently within the construction.  Preliminary results for SU(2) in D\,=\,2+1 at $\beta$=6 are promising.  Section~\ref{sec:jops} reviews the technique for making operators of arbitrary spin J.  Section~\ref{sec:newsmear} describes how to construct superlinks between arbitrary sites using matrix multiplications.  In Subsection~\ref{subsec:tradops} we build traditional operators for Polyakov loops and for spins 0 and 2 glueball operators.  Construction of 8-fold operators is demonstrated in Subsection~\ref{subsec:8fold}, and ``clock'' operators with 12-fold symmetry are in Subsection~\ref{subsec:clockops}.  We highlight the difficulties with higher spin identification in Section~\ref{sec:identification}, and Section~\ref{sec:results} presents results on an L\,=\,16 lattice at $\beta=6$.  Some considerations and suggested improvements follow in Section~\ref{sec:consider}, and we conclude by summarizing.

\section{\label{sec:jops}Construction of Operators with Arbitrary Spin on a Square Lattice}

To resolve the {\it modulo} 4 ambiguity in the spin of conventional operators, one needs to construct operators of arbitrary spin from gauge-invariant loops at relative angles $\theta = 2\pi/n$ other than the $\pi/2$ commonly available on the square lattice.  These operators will suffer from their own {\it modulo} n spin ambiguities determined by the n-fold symmetry employed, but knowledge is power, and control over the value of n lets one select which sector of the spectrum one resolves.  Loops which are $\pi$-symmetric will resolve the positive-parity, even-spin sector, and loops which are not $\pi$-symmetric will be able to resolve the full spectrum.

Consider a timeslice of D\,=\,2+1 in the continuum.  Let $\cC$ be the trace of the gauge field along an arbitrarily shaped closed loop of contour C aligned to an axis.  Copies rotated through an angle $\theta$ are denoted $\cC_\theta$.  Then, to construct a gauge invariant operator $\cO_J$ of arbitrary spin J, we take the weighted average over all loops $\cC_\theta$, where the weight is the complex phase $e^{iJ\theta}$:
\beq
\cO_J = {\frac{1}{2\pi}} \oint d\theta \; e^{iJ\theta} \cC_\theta \;.
\eeq
(The shape and orientation of the loop may influence the phase cancellations.)
On the lattice timeslice, our choice of angles is restricted to the n-fold angles of our chosen operator
\beq
\label{eqn:Jop}
\cO_J \rightarrow {\frac{1}{2\pi}} \sum_{j=1}^n \Delta\theta_j \; e^{iJ\theta_j} \cC_{\theta_j} \;,
\eeq
and our loops $\cC_{\theta_j}$ will no longer be exact rotational copies of each other.  Following \cite{johnson-1999-73}, we normalize all these loops relative to their root-mean-square norm, in order to give the loops approximately equal vacuum expectation values, {\it ie}
\beq
\cC_{\theta_j} \rightarrow \cC_{\theta_j} \times \sqrt{ \langle \langle \cC_{\theta_j}^2 \rangle \rangle_j / \langle \cC_{\theta_j}^2 \rangle } \;,
\eeq
where $\langle \cdots \rangle_j$ denotes the average over angles $j$.  In practice, we find the smaller loops return roughly equal vevs after the above normalization, while the vevs of the larger loops display greater variance.  Vacuum-subtracted correlation functions are defined by
\beq
\label{eqn:corrfcn}
C_J(t) \equiv \langle \cO_J^\dagger(t) | \cO_J(0) \rangle - \langle \cO_J^\dagger \rangle \langle \cO_J \rangle \;,
\eeq
and are normalized to $C_J(0)=1$.  For complex-valued operators we take the real part of Equation~(\ref{eqn:corrfcn}) as our correlation function.  Similarly, we can compute the normalized cross-correlation for states of different J:
\beq
C_{{\widetilde{J}}J}(t) \equiv \frac{\langle \widetilde{\cO}^\dagger_{\widetilde{J}}(t) | \widetilde{\cO}_J(0) \rangle}{\sqrt{\langle \widetilde{\cO}^\dagger_{\widetilde{J}} | \tilde{\cO}_{\widetilde{J}} \rangle \langle \widetilde{\cO}^\dagger_J | \widetilde{\cO}_J \rangle}} \;,
\label{eqn:overlap}
\eeq
where the tilde over $\cO$ implies the vacuum-subtracted operator.  The timeslice overlap between the operators is identified by the value of $t$ in Equation~(\ref{eqn:overlap}).  We restrict ourselves in this article to using loops which are $\pi$-symmetric---if we were to use closed loops which were not symmetric under a rotation by $\pi$, we could recover both even and odd spins and both positive and negative parity, otherwise phase cancellations restrict us to the positive-parity, even-spin sector of the spectrum.

\section{\label{sec:newsmear}Smeared Links from Matrix Multiplication}

To construct our arbitrarily shaped loops, we need to evaluate smeared links between arbitrary sites in the timeslice.  We start by defining the link supermatrix $\cM$, indexed by site, of dimension $L^{D-1} \times L^{D-1}$, whose elements, themselves SU(N) matrices (giving a total matrix dimension of $NL^{D-1} \times NL^{D-1}$), are constructed from the timeslice links $U_{i,j}$ from sites $i$ to sites $j$ as follows:
\beq
\cM_{i,j} \equiv \left\{ \begin{array}{llr} 0 & {\textrm{if sites i and j are not nearest neighbors}} & \\ U_{i,j} & {\textrm{if sites i and j are nearest neighbors}} & \;.
\end{array} \right.
\eeq
Note that $\cM_{j,i} = U_{i,j}^\dagger$.  Previously \cite{johnson-2002-66b,meyer-2003-658}, Green's functions were calculated by noting that
\beq
\cK \equiv \frac{1}{1 - \alpha \cM} \approx 1 + \alpha\cM + \alpha^2\cM^2 + ...
\eeq
for some small parameter $\alpha$ which behaves much like a ``hopping parameter'' for a scalar test particle sampling the gauge field. The matrix $\cM^k$ consists of all paths of length $k$---for even $L$, the odd powers of $\cM$ are traceless while the even powers of $\cM$ have a trace consisting of all paths of length $k$ which terminate at their starting site, including those which double back on themselves.  The diagonal elements of $\cM^2$ are ``unity paths'', 4 for each site, {\it ie} $\mathit{Tr}\cM^2 / 4NL^{D-1} = 1$.  The expectation value of the spatial plaquette $\langle U^S_\Box \rangle$ may thus be written as
\beq
\langle U_\Box^{M^4} \rangle \equiv ( \mathit{Tr}\cM^4 / NL^{D-1} - 28 )/ 8 \;.
\eeq
Note that this formula corrects that given previously \cite{meyer-2003-658,RWJ:2002th}, as there are actually 7 paths which contribute unity per lattice direction and each plaquette is counted 8 times (4 corners by 2 orientations).  Expectation values for larger closed loops may be found from the traces of higher even powers of $\cM$, {\it eg} the vev of the spatial 1x2 rectangle $\langle U^S_{\sqsubset \sqsupset} \rangle$ will relate to the trace of $\cM^6$---these might prove useful if incorporating an improved action~\cite{morningstar-2000-83} into the calculation.

Rather than computing the inverse matrix $\cK$, which suffers from torelon contributions coming from the terms in the infinite series with $k > L/2$, we compute the sequence of matrices $\cM^k$ for $k \in \{1,k_{max}\}$.  (``Torelon contribution'' is actually a bit of a misnomer---the real problem is that for a distance $k > L/2$ between chosen sites $i$ and $j$, $\cM^k_{i,j}$ may contain contributions from both ways around the periodic lattice.  With a suitable reduction in the unity path contributions (below) necessary for a path the short way around to be of distance $k > L/2$, these effects may be ameliorated.)   For $k_{max} > L/2$, torelon contributions are determined by the locations of sites $i$ and $j$---we may propagate up to a total distance of $L$ along a pure diagonal before contamination sets in.  In practice we take $k_{max} = L$ to allow maximum propagation and avoid contamination by judicious choice of superlink orientation when constructing operators, or we may simply remove operators {\it ab finito} which appear to have been contaminated.  Note that our parameter $\alpha$ has disappeared---we may reintroduce a smearing level parameter by applying a number $p$ of smearing iterations, conveniently written as
\beq \label{eqn:smear}
\cM^{p+1} \Leftarrow  [\mathit{mask}(\cM^p) \times (\cM^p)^3 - 5\cM^p]/4 \;,
\eeq
where $\mathit{mask}(\cM^p) \times$ simply removes those elements of $(\cM^p)^3$ which do not correspond to nearest neighbors, before constructing our matrices $\cM^{k, p}$ (at pre-smearing level $p$).  Blocking algorithms may be implemented by a different choice of masking.   We are now almost ready to use our superlinks $\cM^{k, p}_{i,j}$ to construct our gauge-invariant loops.

\begin{figure}
\includegraphics[scale=.5,bb=100 200 500 600]{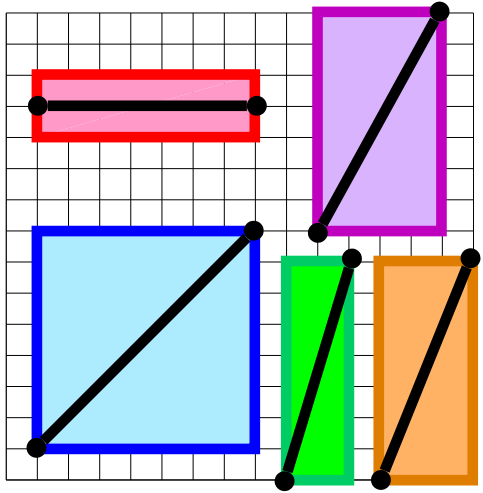}
\caption[Superlinks of various widths and orientations.]{\label{fig:superlinks}(Color online.)  Superlinks of various widths and orientations.  The superlink along the central line contains contributions from all the links covered by the surrounding shaded region.}
\end{figure}

In order to reduce the count of unity paths contributing to our $\cM^k$ (dropping the index $p$), first we form
\beq
\widetilde{\cM}^2 = \cM^2 - diag(\cM^2) \;,
\eeq
where we have removed the diagonal elements of $\cM^2$.  Then we construct our sequence of matrices iteratively:
\bea \label{eqn:makeMs}
\widetilde{\cM}^{2k+1} &=&  \widetilde{\cM}^{2k} \times \cM \; \; , \\
\widetilde{\cM}^{2k+2} &=&  \widetilde{\cM}^{2k} \times \widetilde{\cM}^2 - diag(\widetilde{\cM}^{2k} \times \widetilde{\cM}^2) \; \; ,
\eea
removing the diagonal elements of the even powers as we go.  As these matrices are sparse, performing the multiplications may be executed quite efficiently, and storage requirements are modest---the superlink construction requires but a fraction of the memory needed to compute the cross-correlations of the binned operators.  We may verify our construction by checking that the spatial plaquette equals
\beq
\langle U^{\widetilde{\cM}^4}_\Box \rangle \equiv ( \mathit{Tr}\widetilde{\cM}^4 / NL^{D-1} - 12 )/ 8 \;,
\eeq
indicating a reduction in unity paths from 28 to 12.  If we examine the superlinks along axes directions (Figure~\ref{fig:superlinks}), we see that to propagate a distance $l$ between sites $i$ and $j$ along an axis, we should take $\widetilde{\cM}^{l+2}_{i,j}$, and to propagate a distance $l$ off-axis we use $\widetilde{\cM}^l_{i,j}$, where $l$ is defined to be the number of links in the shortest path between sites $i$ and $j$.

The last step before we can construct our operators is to perform the reunitarization of the superlinks.  The elements of $\cM^k$, consisting of the sums of the possible paths, are not going to be elements of the gauge group.  A preliminary calculation using unreunitarized superlinks projected well onto the ground states in the various sectors, but the excited states were impossible to recover cleanly.  For the case at hand, SU(2), the reunitarization is simple: just divide each SU(2) element by its determinant.  That simple step accounts for more than half the computational time currently consumed.  For general SU(N), the reunitarization involves either an iterative procedure to maximize the trace of the projection or repetitive calculations of eigenvectors to unitarize the elements\cite{Liang:1992cz,meyer-2005}---either way, a lengthy part of the calculation.  Explorations of whether the matrix method for superlinks can return a naturally unitarized smeared link are underway.

\section{\label{sec:ppops}Construction of Positive-parity Operators}

\subsection{\label{subsec:tradops}Construction of Traditional Operators}

\begin{figure}
\includegraphics[scale=.5,bb=100 200 500 600]{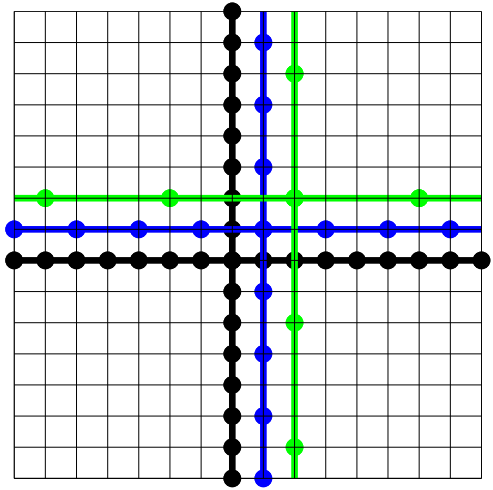}
\caption{\label{fig:poly}(Color online.)  Operators for Polyakov loops at sizes 1, 2, and 4.}
\end{figure}

\begin{figure}
\includegraphics[scale=.5,bb=100 200 500 600]{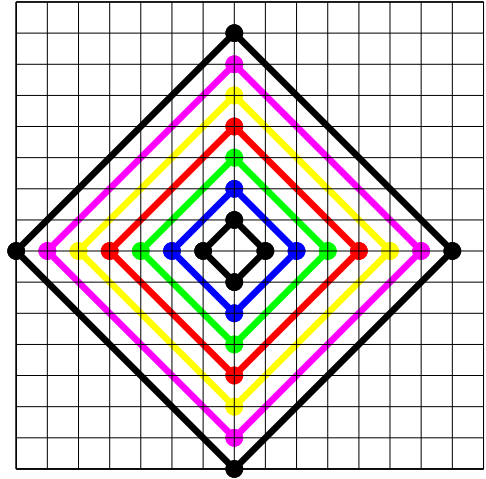}
\caption{\label{fig:box}(Color online.)  Box operators at sizes 1, 2, 3, 4, 5, 6, and 7.}
\end{figure}

\begin{figure}
\includegraphics[scale=.5,bb=100 200 500 600]{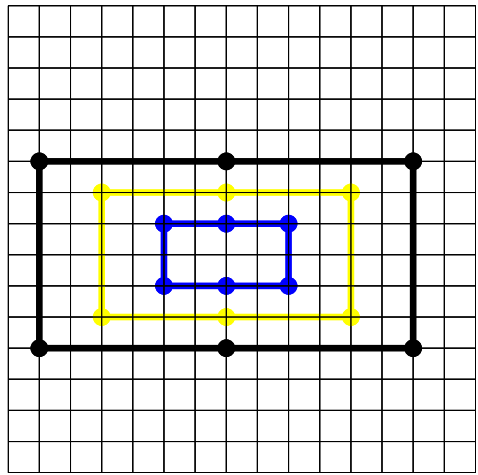}
\caption{\label{fig:bar}(Color online.)  Bar operators at sizes 2, 5, and 7.}
\end{figure}

Operators for Polyakov loops may be constructed in $D-1$ directions, at sizes $l$ for which $L/l$ is an integer, simply by taking the product of superlinks along an axis $\hat{\mu}$:
\beq
\cO_{P, \; \hat{\mu}}^l = \prod_{m=0}^{L/l-1} \widetilde{\cM}^{l+2}_{i+ml\hat{\mu}, \; i+(m+1)l\hat{\mu}} \;,
\eeq
see Figure~\ref{fig:poly}.  ``Box'' operators of size $l$ centered at site $i$ may be formed as
\beq
\cO_\Box^l = \widetilde{\cM}^l_{i+l\hat{\mu}, \; i+l\hat{\nu}} \times \widetilde{\cM}^l_{i+l\hat{\nu}, \; i-l\hat{\mu}} \times \widetilde{\cM}^l_{i-l\hat{\mu}, \; i-l\hat{\nu}} \times \widetilde{\cM}^l_{i-l\hat{\nu}, \; i+l\hat{\mu}} \;,
\eeq
where $\hat{\mu},\hat{\nu}$ are orthogonal directions, as in Figure~\ref{fig:box}.  ``Bar'' operators, longer along the $\hat{\mu}$ axis than the $\hat{\nu}$, may be built using the clock points defined below as
\bea
\cO_{B, \; \hat{\mu}}^l &=& \widetilde{\cM}_{X4_l, \; Xm1_l} \times \widetilde{\cM}_{Xm1_l, \; X2_l} \times\widetilde{\cM}_{X2_l, \; X10_l} \nonumber \\[-3mm]
 & & \\[-3mm]
 & & \times \; \widetilde{\cM}_{X10_l, \; Xm2_l} \times \widetilde{\cM}_{Xm2_l, \; X8_l} \times\widetilde{\cM}_{X8_l, \; X4_l} \nonumber \;,
\eea
where $Xm1_l$ is the on-axis midpoint between $X2_l$ and $X4_l$, $Xm2_l$ is between $X2_l$ and $X4_l$, and similarly for $\cO_{B, \; \hat{\nu}}^l$, Figure~\ref{fig:bar}.  Operators of spin 0 and 2 are constructed from the bar operators as the normalized sum and difference of $\cO_{B, \; \hat{\mu}}^l$ and $\cO_{B, \; \hat{\nu}}^l$, {\it ie}
\beq
\cO_{n, \; J}^l = \cO_{4, \; 0/2}^l = ( \cO_{B, \; \hat{\mu}}^l \pm \cO_{B, \; \hat{\nu}}^l ) / 2 \;.
\eeq
Note that these operators are $\pi$-symmetric, 4-fold operators, hence the subscript 4.  Here is the first example of when our construction algorithm returns trial loops of which some are highly redundant, sharing more than one superlink in common with loops of a different size.  Inspection to remove redundant loops is best done graphically, giving easy identification of which loops are not redundant.  States with spins $|J|$ congruent {\it modulo} 4 will have equivalent phases and hence the greatest mutual overlap, {\it ie} spins 0/4,~2/6, and 1/3 if we were using traditional L-shaped loops, will be mixed.  The state with the lowest spin might not be the lightest state in a channel, thus care must be taken when assigning spin quantum numbers to the calculated states.  From established results\cite{teper-1999-59,meyer-2003-658}, we expect the gap between the ground states for the $0^+$ and $4^+$, and between the $2^+$ and $6^+$, to be large enough that the lower spin dominates the lightest few states.

\subsection{\label{subsec:8fold}Construction of 8-fold Operators}

\begin{figure}
\includegraphics[scale=.5,bb=100 200 500 600]{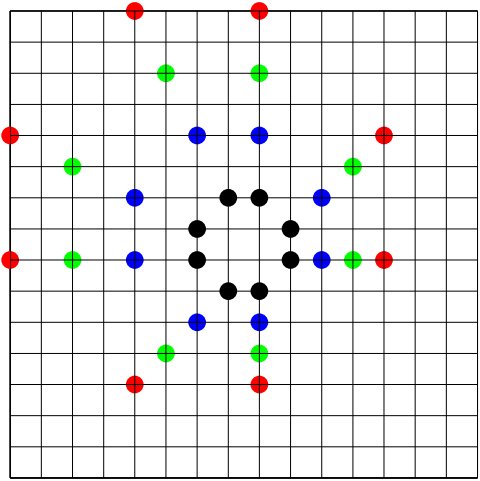}
\caption{\label{fig:8pnts}(Color online.)  8-fold points at sizes 1, 2, 3, and 4.}
\end{figure}

\begin{figure}
\includegraphics[scale=.25,bb=100 200 500 600]{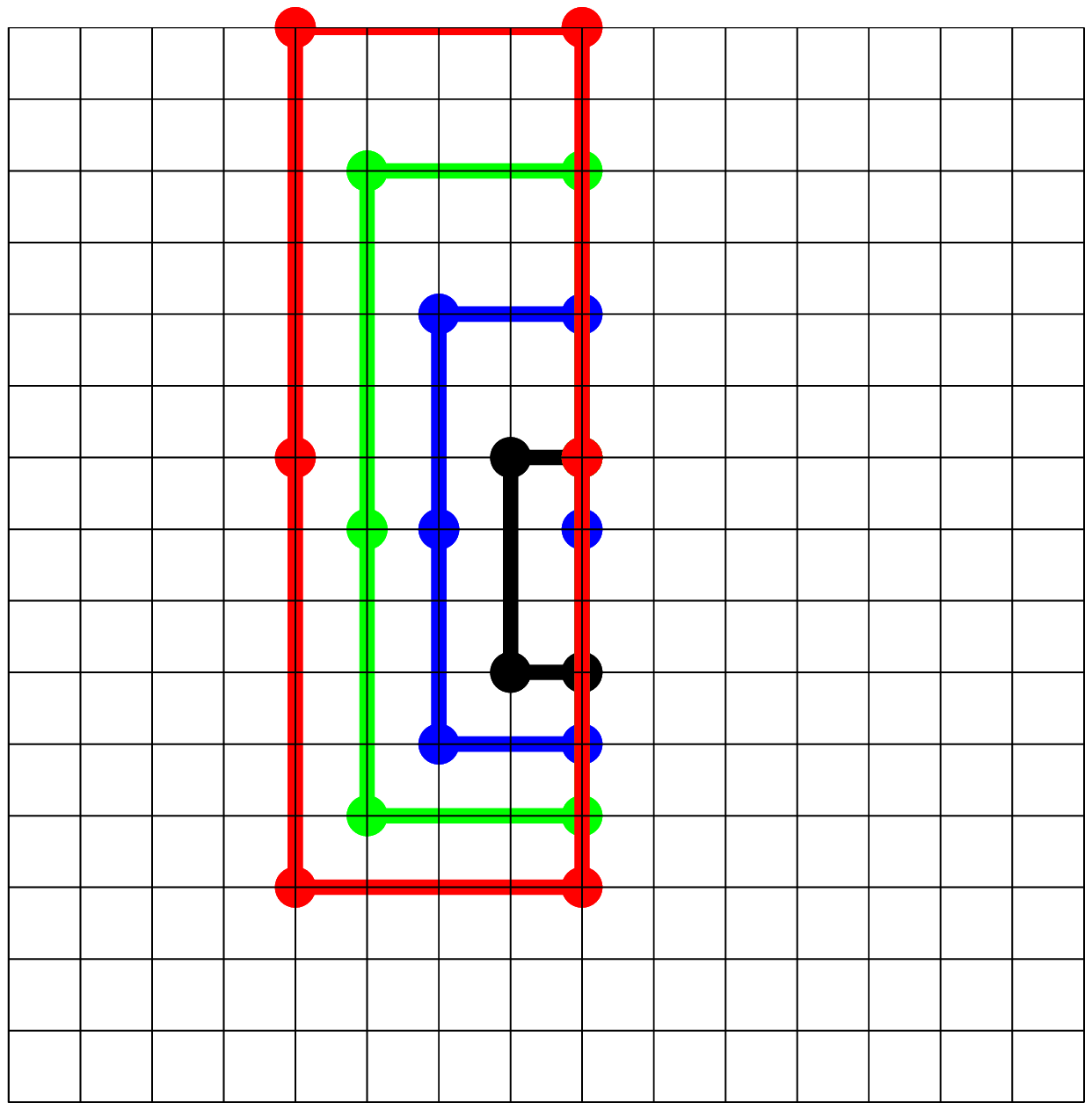}%
\hspace{1in}%
\includegraphics[scale=.25,bb=100 200 500 600]{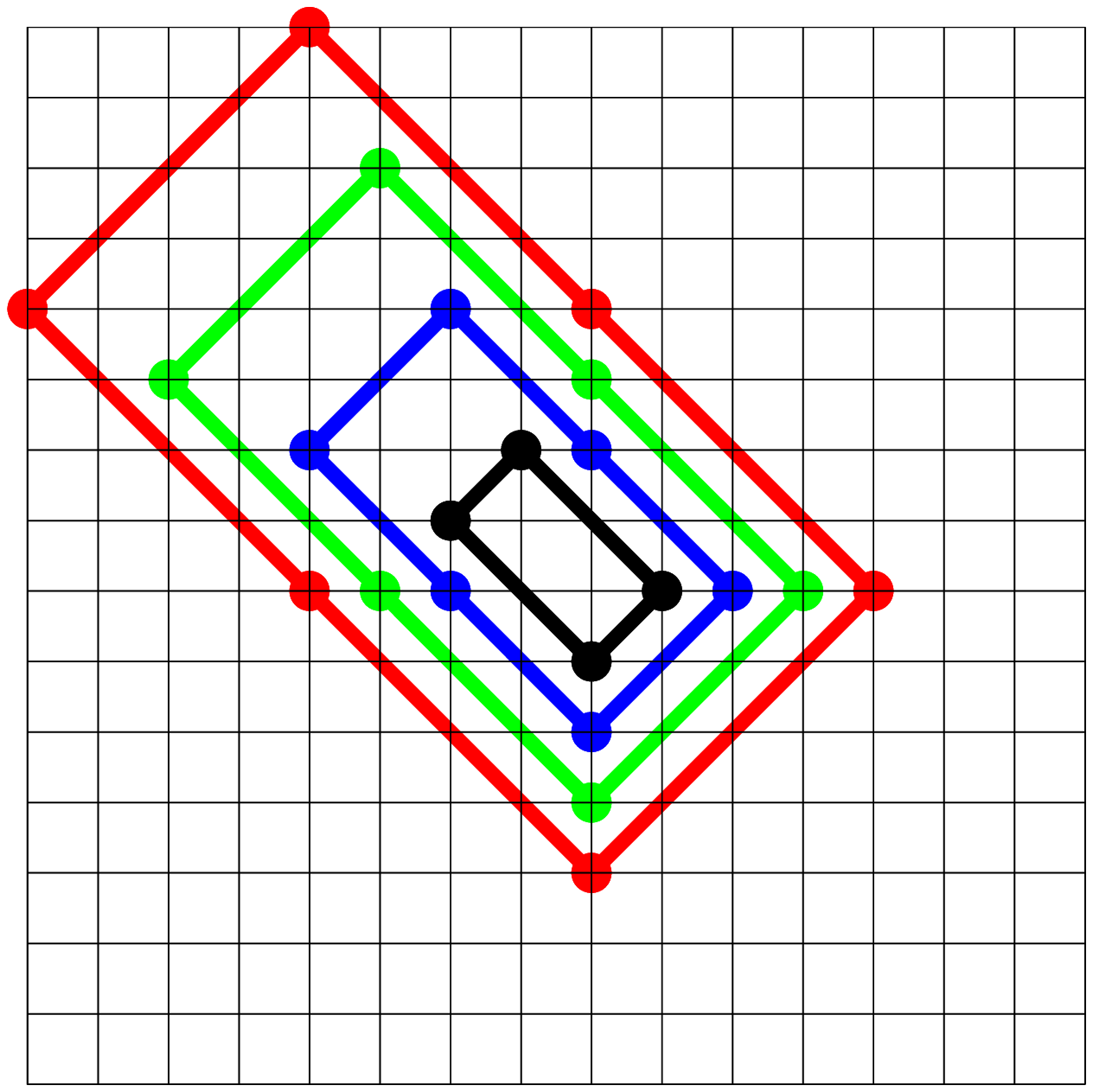} \\
\includegraphics[scale=.25,bb=100 200 500 600]{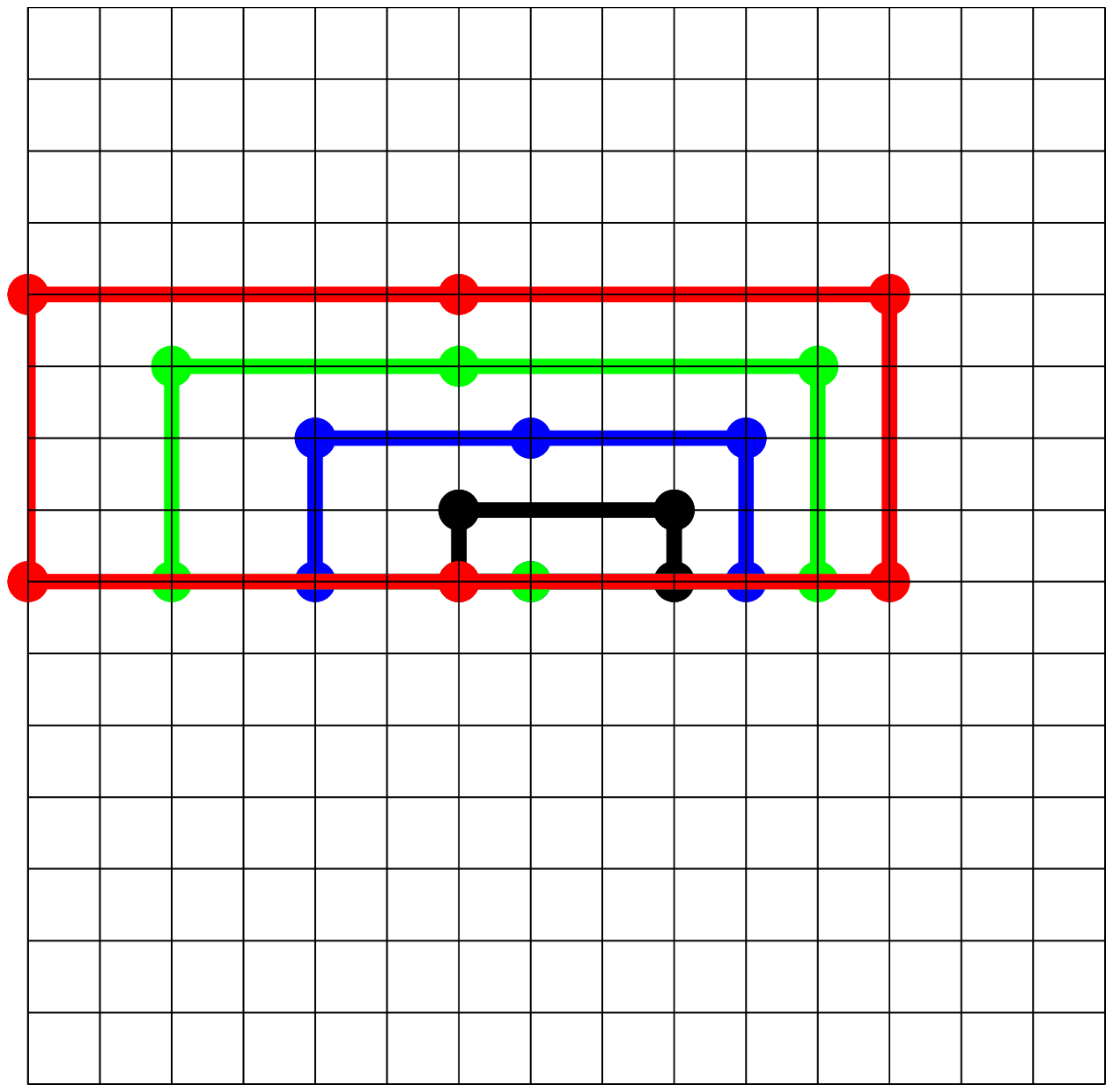}%
\hspace{1in}%
\includegraphics[scale=.25,bb=100 200 500 600]{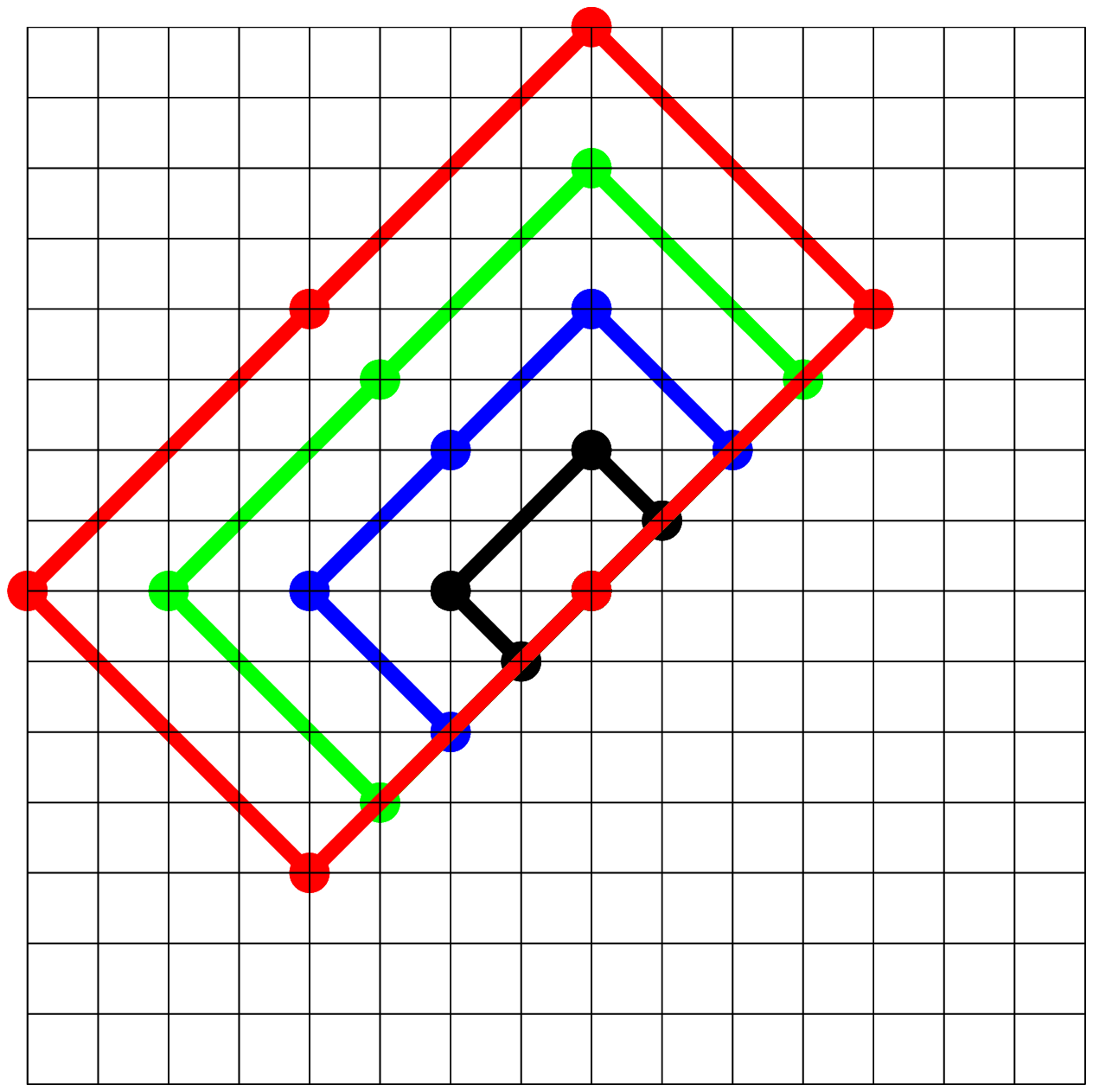}
\caption{\label{fig:8bars}(Color online.)  8-fold operators at sizes 1, 2, 3, and 4.}
\end{figure}

To achieve a degree of symmetry greater than the traditional 4-fold operators, we develop an algorithm to select points in a timeslice which are at relative angles of approximately $2\pi/8$ by placing an octagon on the lattice and labeling the points at size $l$ as $A_l$ through $H_l$ clockwise, Figure~\ref{fig:8pnts}. Using these we may construct 8-fold $\pi$-symmetric operators, {\it eg}
\beq
\cO_8^l = \widetilde{\cM}_{B_l, \; A_l}^d \times \widetilde{\cM}_{A_l, \; F_l}^d \times\widetilde{\cM}_{F_l, \; E_l}^d \times\widetilde{\cM}_{E_l, \; B_l}^d \;,
\eeq
with $d=d(l)$ chosen appropriately, Figure~\ref{fig:8bars}, allowing us to resolve the spin 0/4 ambiguity.  In practice, midpoints are also defined (as for the bar operators of the previous subsection) so that contamination is avoided.  To build operators $\cO_{8, \; J}^l$ with even spin J, we take the weighted linear combinations as above, Equation~(\ref{eqn:Jop}).  States with spins $|J|$ congruent {\it modulo} 8 will have equivalent phases and hence the greatest mutual overlap, {\it ie} spins 0/8,~2/6, and 4/12 will be mixed, but the overlap between the spin 0 and spin 4 should be minimal.  In practice, we find that the trial $4^+$ operator is more than capable of coupling to $0^+$ states, and much care must be taken to distinguish a good $4^+$ from the other trial operators.  The same applies to the $2/6$ sector.  As these operators are themselves ``bar'' operators, we may also form a pair of $2^+$ operators, $\cO_{8-bar, \; 2}^l$, from appropriate pairings of the $\cO_8^l$.

\subsection{\label{subsec:clockops}Construction of Clock Operators}

\begin{figure}
\includegraphics[scale=.5,bb=100 200 500 600]{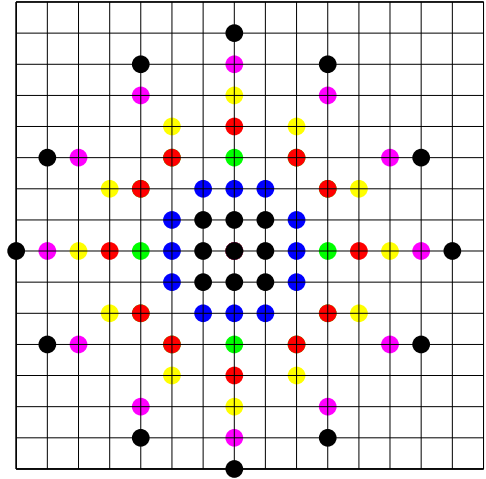}
\caption[12-fold clock points at sizes 1, 2, 3, 4, 5, 6, and 7.]{\label{fig:clkpnts}(Color online.)  12-fold clock points at sizes 1, 2, 3, 4, 5, 6, and 7.  Note the redundancy in corner points for size 1 and between sizes 3 and 4.}
\end{figure}

\begin{figure}
\includegraphics[scale=.25,bb=100 200 500 600]{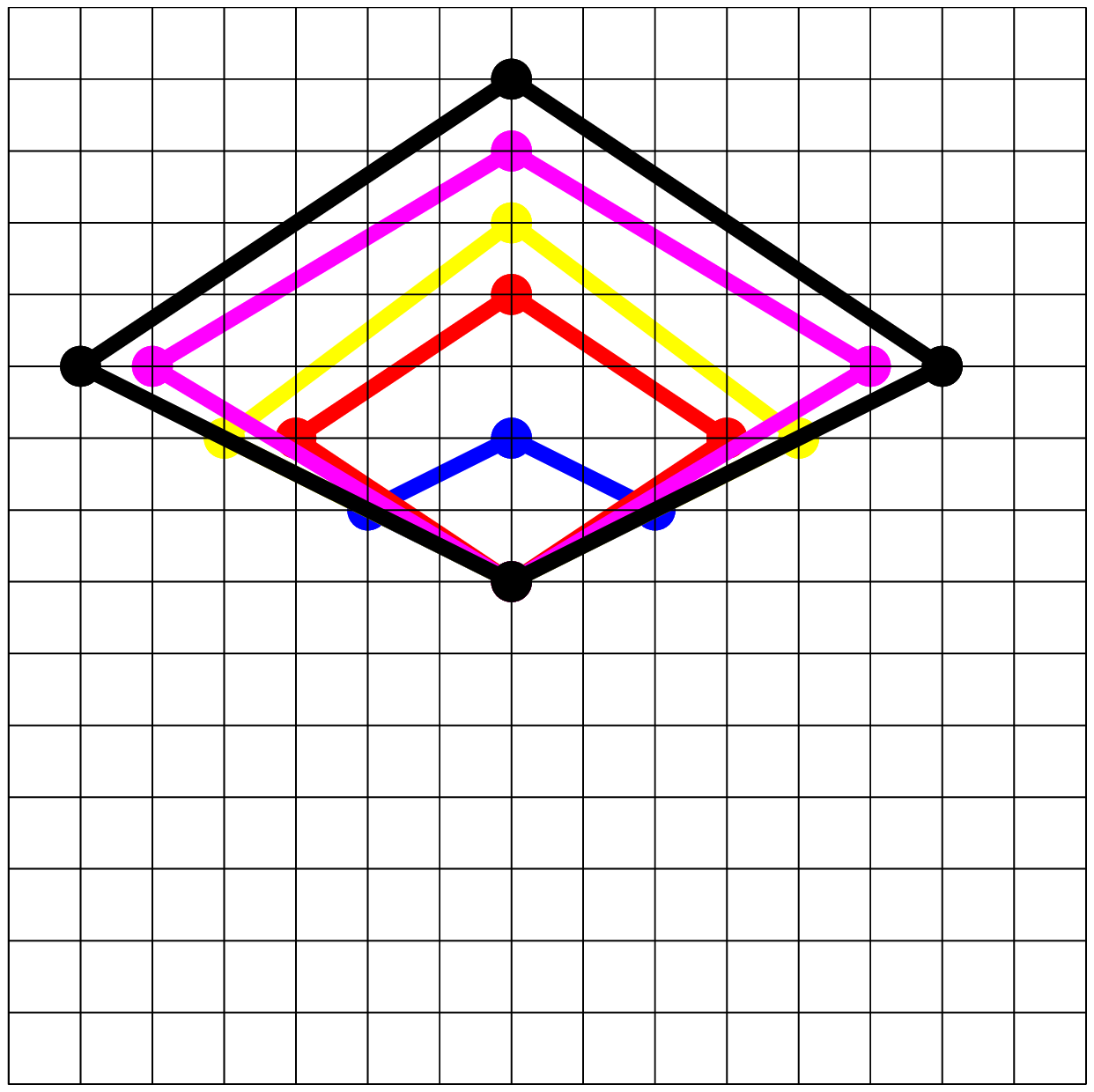}%
\hspace{1in}%
\includegraphics[scale=.25,bb=100 200 500 600]{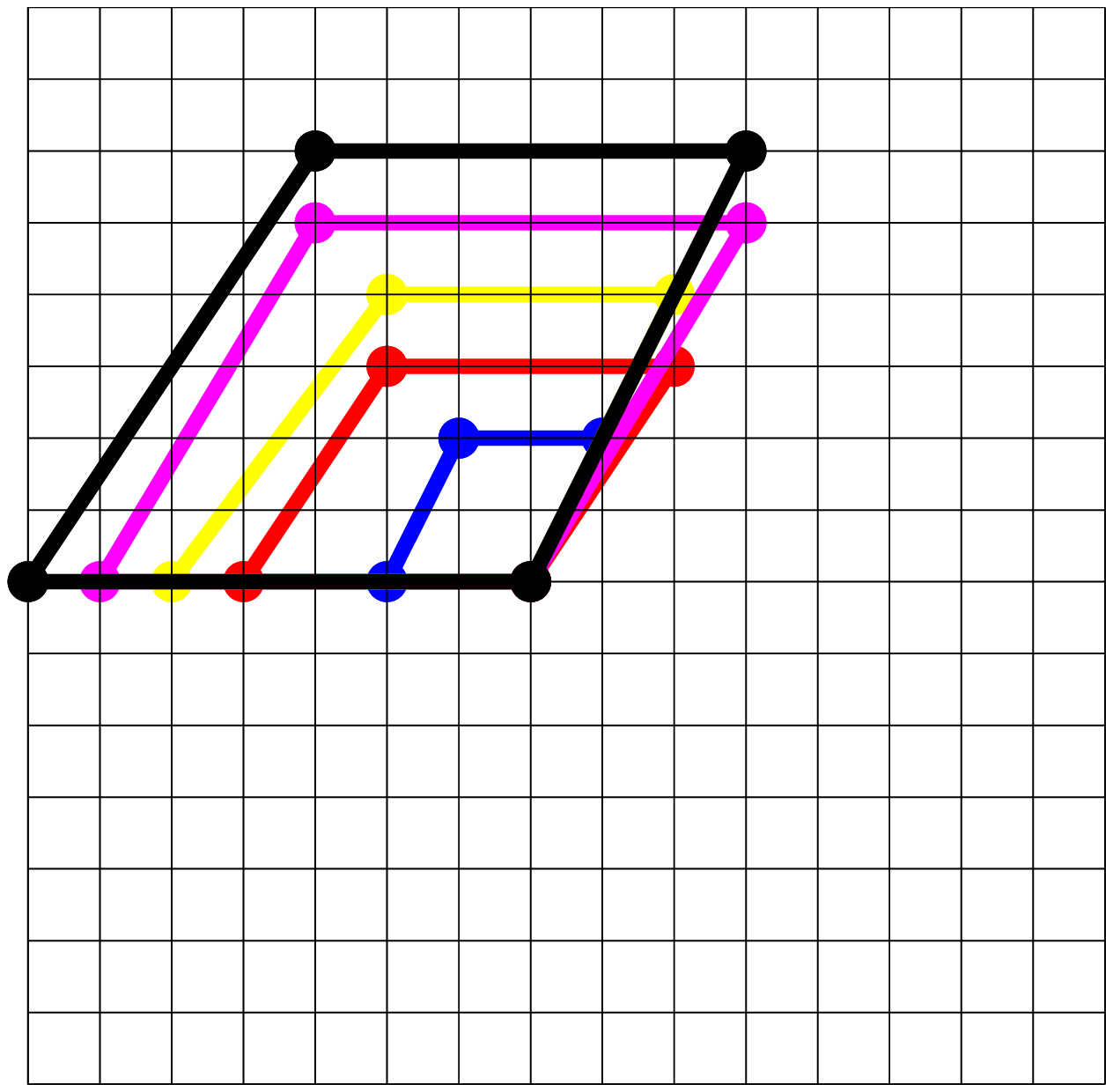} \\
\includegraphics[scale=.25,bb=100 200 500 600]{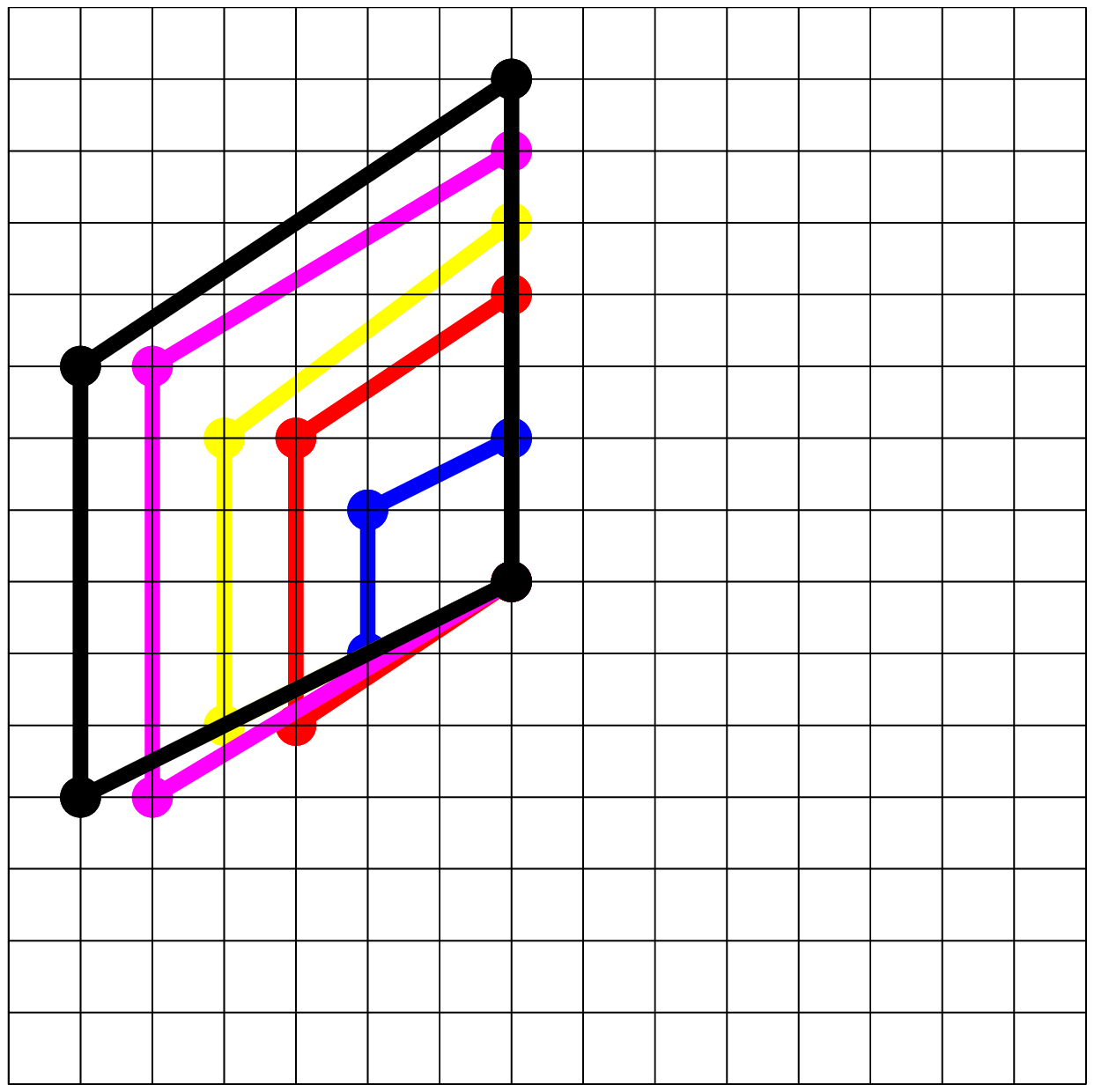}%
\hspace{1in}%
\includegraphics[scale=.25,bb=100 200 500 600]{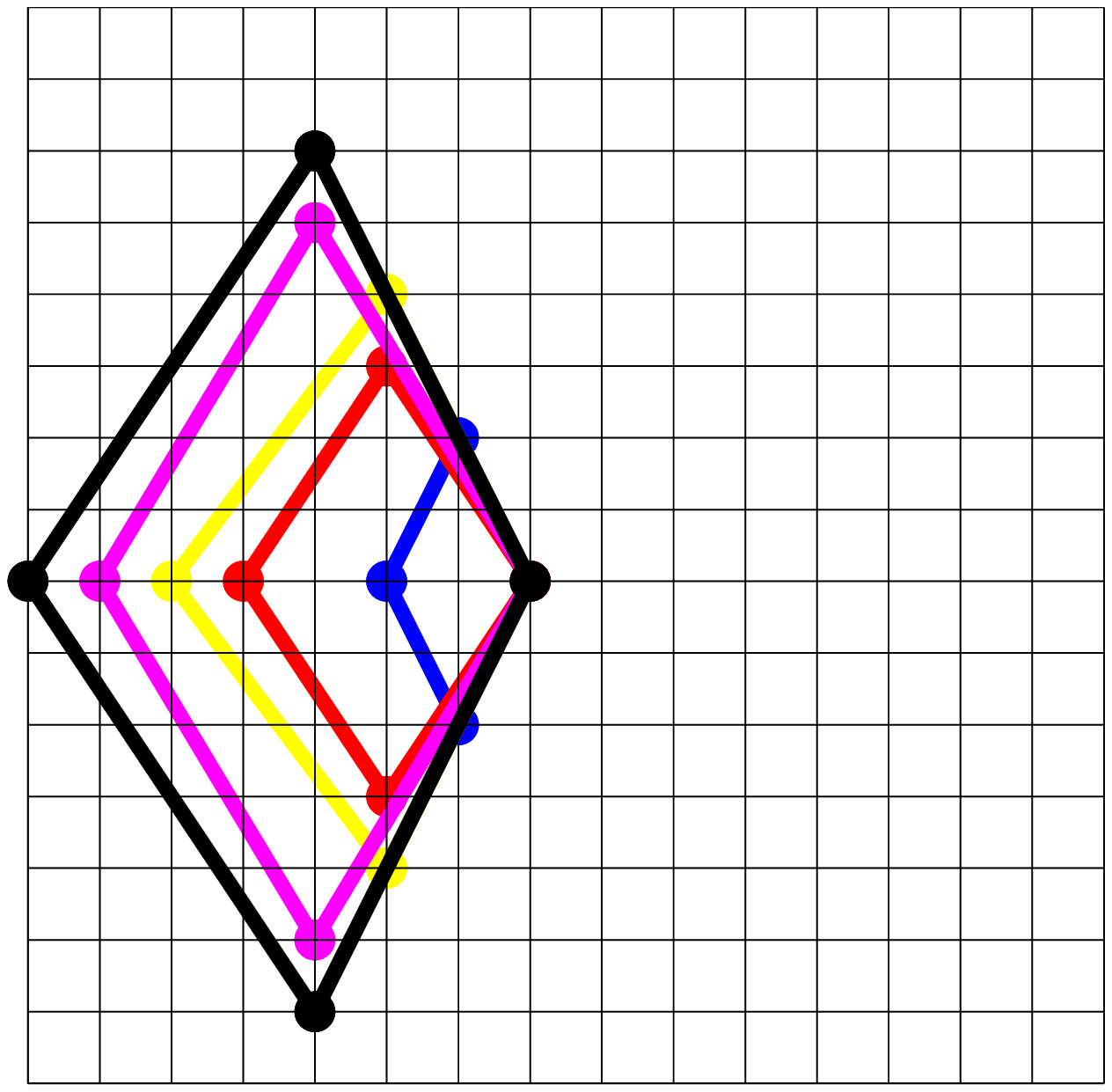} \\
\includegraphics[scale=.25,bb=100 200 500 600]{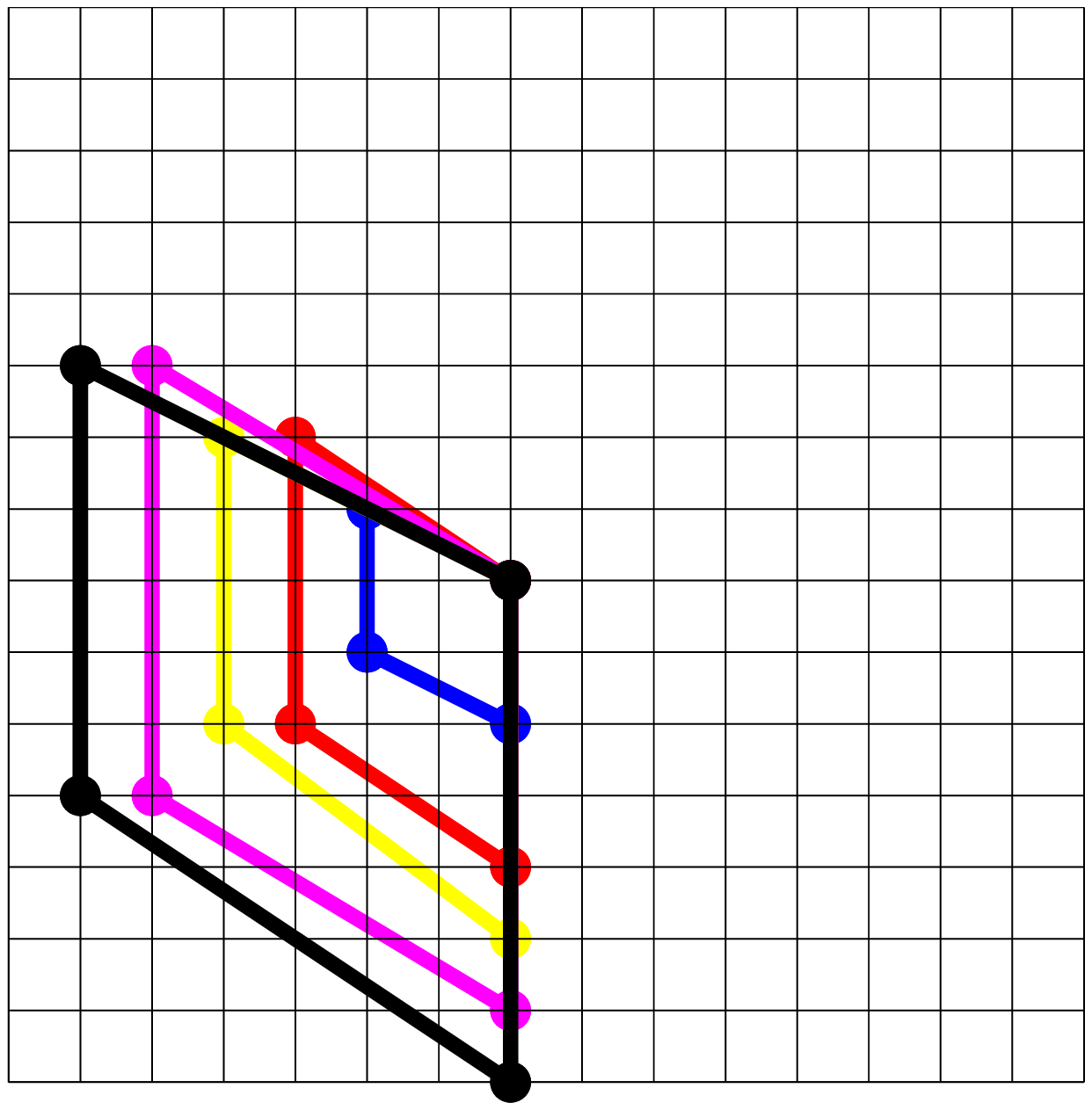}%
\hspace{1in}%
\includegraphics[scale=.25,bb=100 200 500 600]{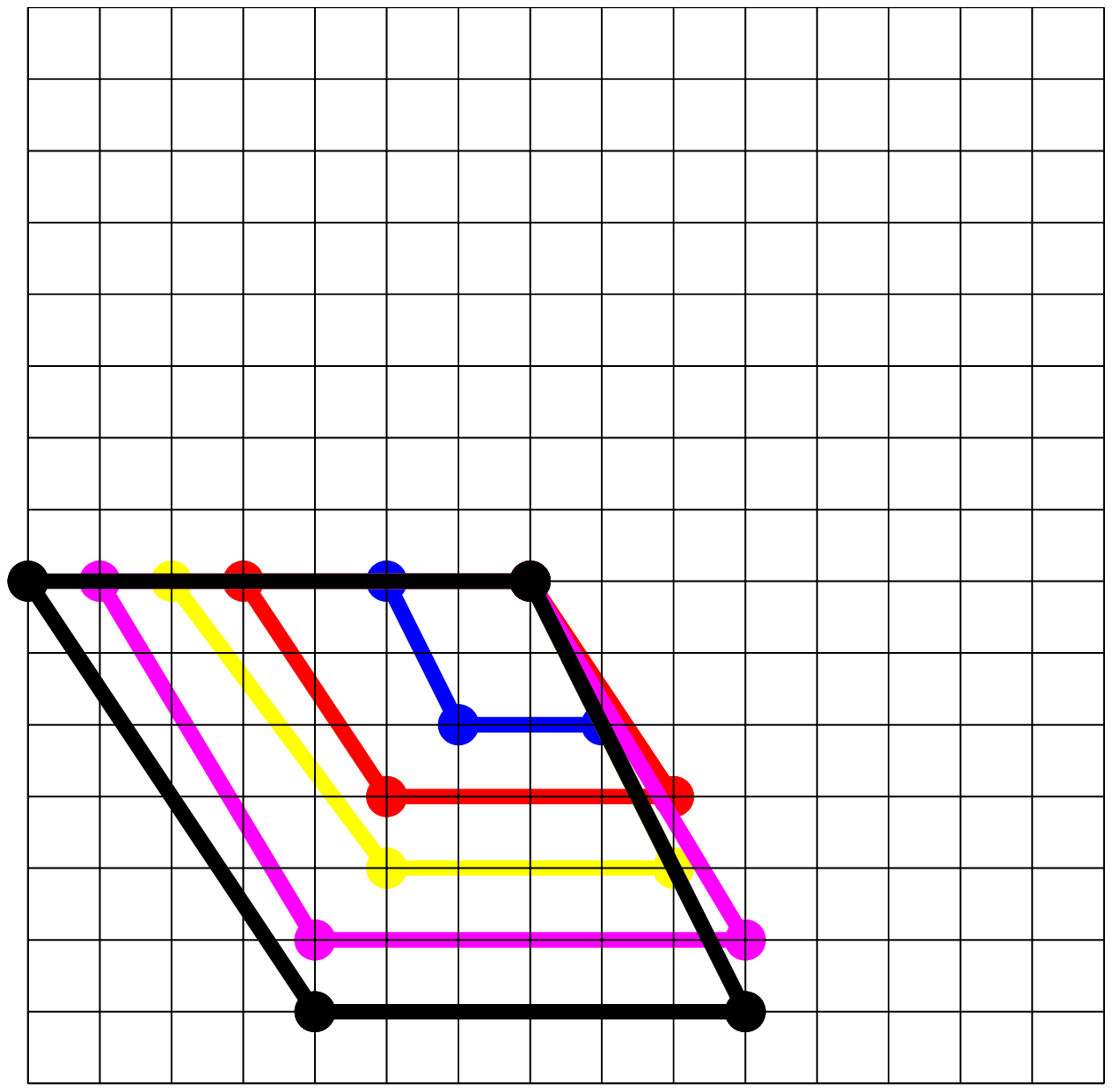}
\caption{\label{fig:rhomops}(Color online.)  12-fold rhomboid operators at sizes 2, 4, 5, 6, and 7.}
\end{figure}

As before, we develop an algorithm to select points in a timeslice which are at relative angles of approximately $2\pi/12$; {\it ie}, we place an imaginary clock face on the lattice centered at site $X$ with increasing sizes and select those sites closest to where the hour marks would be, hence the appellation ``clock''' to operators made from this set of points, Figure~\ref{fig:clkpnts}.  We label these points at size $l$ as  $1_l$ through $12_l$, clockwise naturally.  We define 6 rhomboid shaped loops at angles $\theta_j =j\pi/6$ for $j \in \{0,\ldots,5\}$, {\it eg}
\beq
\cO_{\lozenge j}^l = \widetilde{\cM}_{X, \; 10_l}^d \times \widetilde{\cM}_{10_l, \; 12_l}^d \times \widetilde{\cM}_{12_l, \; 2_l}^d \times \widetilde{\cM}_{2_l, \; X}^d \;,
\eeq
as in Figure~\ref{fig:rhomops}.  To build operators $\cO_{12, \; J}^l$ with even spin J, we take the weighted linear combinations as above.  States with spins $|J|$ congruent {\it modulo} 12 will have equivalent phases and hence the greatest mutual overlap, {\it ie} spins 0/12,~2/10,~4/8, and 6/18 will be mixed.

\section{\label{sec:identification}Identification of Higher Spin Operators}

The operators constructed above for the higher spins 4 and 6 are only exact in the continuum---on a finite lattice Hamiltonian eigenstates must transform as an irreducible representation of the square group $C_{4v}$.  Thus, our trial operators for both $0^+$ and $4^+$ states will belong to the lattice irrep $A_1$, and our trial $2^+$ and $6^+$ will belong to $A_3$.  Using superlinks smeared over some width to construct the loops, the phase cancellations intended to be angular may in fact be radial, resulting, {\it eg}, in a trial $4^+$ operator coupling to an excited $0^+$ state.  As has been known for some time \cite{Johnson:1982yq}, ``ascribing a definite spin to a single lattice state is therefore not possible without further dynamical information'' coming from the behavior of the state as the continuum is approached.

Meyer and Teper \cite{meyer-2003-658} present a couple of methods to identify accurately the spin of a calculated Hamiltonian eigenstate, and the interested reader is directed there.  As the focus of this article is on the improved superlink construction and not on the precise identification of the higher spin operators (which requires the continuum limit to be taken), we perform a heuristic spin identification by combining the methods mentioned above.  After the variational procedure has orthogonalized the candidate states in channels $J$ and $\tilde{J}=J+4$, timeslice overlaps are computed via Equation~(\ref{eqn:overlap}) and the $\tilde{J}$ states with the least overlap are identified as such.  Consequently, the masses presented here for the higher spins 4 and 6 should be viewed as preliminary until a continuum limit calculation using the superlink operators has been performed.

\section{\label{sec:results}Results}

\subsection{\label{subsec:calcdtls}Calculation Details}

To evaluate the performance of our superlink construction, we calculate the above operators for pure gauge SU(2) on an L\,=\,16 lattice at $\beta=6$ with 15,000 measurements taken once every 10 compound sweeps, both with ($p=1$) and without ($p=0$) pre-smearing.  Thermal updates are done via the Kennedy-Pendleton heat bath algorithm \cite{Kennedy:1985nu} augmented with a 4:1 ratio of over-relaxation sweeps \cite{Creutz:1987xi} and global gauge transformations every 19 sweeps.  After selecting a set of operators based on their auto-correlation functions, cross-correlations are computed and used in a variational procedure to extract the lightest few states in each channel \cite{teper-1999-59}.  Timeslice overlaps are computed to check the performance of the variational procedure as well as the spin decomposition of the final states.

Zero-momentum states are constructed by averaging operators over the timeslice.  Effective masses are extracted from the normalized correlation functions for $t \geq a$ via the formula
\beq
a m_J^{\mathrm{eff}}(t) = \log\left(\frac{C_J(t-a)}{C_J(t)}\right) \;.
\eeq
In principle, one should include a term representing correlations the other way around the periodic lattice, but in practice such effects are swamped by the statistical noise for lattices with extent $L_t$ great enough, especially for the heavier states.  Errors are estimated using the jackknife procedure~\cite{Montvay:1994cy}.

Standard practice is to observe the effective masses for evidence of a ``mass plateau'', which is taken to be the mass of the lightest state contributing to that correlation function over that region in $t$.  While the variational procedure mentioned above does indeed produce operators which are in some sense orthogonal, there are certainly still contributions from two states (or more) for many of the correlation functions.  In order to extract the most information possible from our correlation functions, we also perform a fit~\cite{fsolv1-1999, fsolv2-1988} to the sum of two exponentials
\beq
C_J(t) \sim A^1 \exp(-m^1_J t) + A^2 \exp(-m^2_J t) \;,
\eeq
for a range of $t \in \{0, 4\}$, a notoriously hard problem.  That difficulty shows up in a couple of ways:  first, the 95\% confidence limits on the parameters returned by the fitting routine are often quite large, and secondly, when unconstrained, some correlation functions return fitted parameters that are essentially nonsensical for the given application.  The first we can circumvent by using the familiar jackknife procedure, the second by constraining the parameters to be physically significant positive quantities.  To accurately calculate the binned values, care must be taken that the binned fits are sorted to match the order of states returned by fitting the ensemble correlation function.  These two approaches to extracting the masses should be seen as complementary.

\subsection{\label{subsec:calcspec}Calculated Spectra}

In Table~\ref{tab:plaqs} we compare the vacuum expectation values for the lattice plaquette and spatial plaquette computed directly and the spatial plaquette computed via $\cM^4$ and $\widetilde{\cM}^4$.  Our directly computed plaquettes are well within errors of the accepted value \cite{teper-1999-59}, and our superlink plaquettes agree perfectly with the direct spatial plaquette.  For the remaining operators, we perform the fits as above.  We will examine the full table of fitted parameters and effective masses for the Polyakov loops and the $0^+$ box operators to demonstrate and evaluate the effectiveness of the fitting procedure---for the remaining operators we will display only the best-fit masses and errors.  Sometimes the correlation function is best fit by a single exponential; that condition shows up as equal amplitudes of 1/2 for the two states.  Otherwise, the state with the larger amplitude is taken as the dominant state in the correlation function.  Note that the sum of the amplitudes was not constrained to unity---that resulted from the fitting procedure itself.

Table~\ref{tab:poly} displays the best fit parameters and the effective masses in lattice units for the Polyakov loops after applying the variational procedure.  Effective masses which were negative or imaginary have been zeroed out in the table.  Jackknife errors are given in parentheses after the mass values.  The apparent four-fold degeneracy is not unexpected, as we have included both orthogonal timeslice directions as well as two pre-smearing levels in the variational procedure.  We highlight the ground state and also what appears to be an orthogonal excited state in boldface.  The ground state Polyakov loop agrees well with the accepted value \cite{teper-1999-59}.

The $0^+$ box operator's parameters and masses are shown in Table~\ref{tab:box}.  We note that now the degeneracy is two-fold, apparently a result of combining the two pre-smearing levels of operators in one variational procedure.  Selected states are highlighted in boldface---preference is given to that state of a pair with either (or both) a single exponential or a lower mass, except for the heaviest, which was chosen for its lower error.  The ground and first excited states compare favorably with previous results \cite{meyer-2003-658}.

Satisfied that the fitting procedure is working reasonably well, for the remaining operators we present only the best fit parameters of the dominant state.  The $0^+$ and $2^+$ bar operators are in Table~\ref{tab:bar02}, and the $2^+$ 8-bar operators of orthogonal orientation are in Table~\ref{tab:8bar2}.  These masses compare well with the values given by Teper and Meyer and are seen to be self-consistent.  That the $2^+$ states are best fit by masses slightly higher than conventional seems to result from the conventional act of selecting the second effective mass as the best estimate, which is, after all, a highly arbitrary procedure.  The $0^+$, $2^+$, and $4^+$ best fit masses for the 8-fold operators are shown in Table~\ref{tab:8fold024}.  For the $4^+$ we have selected the state with the least overlap with the $0^+$ (see Table~\ref{tab:8foldovrlaps}), which is in fair agreement with the value by Meyer, excluding the extremely heavy state, but we should keep an open mind as to the identification, noticing a degeneracy with an excited $0^+$ in the same table.  The $0^+$, $2^+$, $4^+$, and $6^+$ best fit masses for the 12-fold operators are shown in Table~\ref{tab:12fold0246}.  The $0^+$ displays and odd splitting of the ground state, and we highlight both values.  The excited $2^+$ states are somewhat different in mass from the other operators' estimates.  The highlighted ground states for the $4^+$ and $6^+$ are selected on the basis of their overlaps with the $0^+$ and $2^+$, respectively, Tables~\ref{tab:12fold04} and \ref{tab:12fold26}.  (Overlaps between the 0/4 and 2/6 sectors are minimal.)  The masses for the $4^+$ and $6^+$ compare well with those found by Meyer~\cite{meyer-2005}.

\section{\label{sec:consider}Considerations and Improvements}

The use of a pre-smearing level in this context serves more to remove ultraviolet fluctuations from the lattice configuration than to control how ``wide'' the superlinks are, and a better approach to controlling the effective width at various orientations should be developed.  Returning to the technique of removing the diagonal elements of the even powers of $\cM$ in Equation~(\ref{eqn:makeMs}), the relative weightings of the more directional paths with unity path contributions might serve better to control how directional the superlinks are.  As currently defined, the superlinks in Figure~\ref{fig:superlinks} get effectively wider as their orientation approaches the diagonal:  superlinks along an axis span the narrowest rectangle, while superlinks along a pure diagonal span the full square of links in-between.  These exceptionally wide superlinks might explain why radial rather than angular phase cancellations seem to dominate the higher spin operators.  Retaining the unity path contributions would give more weight to the more direct paths along the diagonal.  One might also consider using superlinks of length $l+2$ rather than the minimum $l$ used above---such superlinks would be proportionately wider and might be necessary as the lattice spacing decreases to keep the operators on a physical scale.

The remaining {\it modulo} 4 spin mixing results from the mass eigenstates on a finite 2+1 dimensional lattice properly belonging to irreducible representations of the square group, so our trial operators are going to contain contributions from multiple irreps.  However, in the continuum limit, as the lattice spacing $a$ approaches zero, our trial operators should couple to states of pure spin and mass.  As computer power increases, simulations closer to the continuum limit become possible, and the ability to create operators with arbitrary rotational symmetries will prove useful.

Including both pre-smearing levels in a single variational procedure seems to have introduced a two-fold degeneracy for most of the operators, and in hindsight should probably have been avoided.  Intermediate inspection of the correlation functions with and without pre-smearing indicated a better projection onto the ground states for the pre-smeared operators, so some level of pre-smearing should probably be retained in future calculations.

A direct comparison of the performance of this technique with more traditional smearing techniques has yet to be done, as well as investigating how the performance scales with dimension D and gauge group N.  Such comparisons are forthcoming.  The negative parity and odd spin sectors need to be explored with appropriately shaped operators before the true utility of matrix superlinks can be determined.  Relating these superlinks to those constructed previously \cite{johnson-2002-66b,meyer-2003-658} by calculating the effective $\alpha$ of the current truncated expansion might shed light on both matrix propagator methods as well as on a new angle of approach to simulating staggered fermions on the lattice.

Finally, we note that the superlink matrix $\cM$ need not be confined to a single timeslice---one may form a super-superlink matrix $\cM^D$ from all the links of the given lattice configuration.  By dimensional reduction, the super-superlink matrix in D dimensions corresponds to the timeslice superlink matrix in D\,+\,1 dimensions.  Essentially, $\cM^D$ holds all the information of the lattice configuration in a compact and convenient notation, and the traces of its even powers may be related to the ensemble plaquette average as well as terms used in various improved actions~\cite{morningstar-2000-83}.  In light of Equation~(\ref{eqn:smear}), familiar iterative algorithms may be expressed (and calculated) using efficient matrix notation, and with a suitable modification of the updating algorithm, a single matrix structure could be used for both updating the lattice configuration and calculating the operators.

\section{\label{sec:conc}Conclusion}
The application of the matrix method of superlink construction has been demonstrated to agree with traditionally smeared operators and with previous methods of superlink construction.  The primary difficulties with the matrix method have been alleviated by direct matrix multiplication rather than matrix inversion.  The even spin, positive parity spectrum for $\beta=6$, SU(2) in D\,=\,2+1 dimensions displays excellent agreement with accepted values up to spin 6.  The use of operators less symmetric than the ones presented here would resolve the odd spin and negative parity sectors.  Further exploration of the matrix method of superlink construction is warranted, both to resolve the complete spin and parity spectrum for SU(N) in higher dimensions and potentially to evaluate fermion determinants in SU(N) with quarks.

\section*{\label{sec:ackn}Acknowledgments}
The author gratefully acknowledge continuing conversations with Mike Teper of Oxford University on a variety of topics in lattice gauge theory, as well as computer time provided by Cassiano de Oliveira of the Georgia Institute of Technology.

\newpage

\bibliographystyle{apsrev}

\clearpage

\begin{table}
\centering
\begin{tabular}{|c|c|c|c|} \hline
\multicolumn{4}{|c|}{Plaquette Values} \\\hline
$\langle U_\Box \rangle$ & $\langle U^S_\Box \rangle$ & $\langle U^{\cM^4}_\Box \rangle$ & $\langle U^{\widetilde{\cM}^4}_\Box \rangle$ \\\hline
      0.8248964(11)   &   0.8245756(15)   &   0.8245756(15)   &   0.8245756(15) \\\hline
\end{tabular}
\caption[Plaquette values.]{\label{tab:plaqs}Plaquette values computed directly and with superlinks.}
\end{table}

\begin{table}
\centering
\begin{tabular}{|cccc|cccc|} \hline
\multicolumn{8}{|c|}{Polyakov Loops} \\\hline
\multicolumn{4}{|c|}{Best Fit Parameters} & \multicolumn{4}{c|}{Effective Masses} \\\hline
 Amp1  &    Mass1  &  Amp2  &      Mass2  &   EffM1  &   EffM2  &    EffM3   &    EffM4    \\\hline
0.689 &  {\bf 0.994(017)}  &  0.311 &  2.011(142)   &  1.216(1)  &  1.086(02)  &  1.028(05)  &  1.074(15)  \\
0.858 &  1.077(014)  &  0.142 &  3.685(537)   &  1.221(1)  &  1.065(02)  &  1.077(06)  &  0.947(13)  \\
0.930 &  {\bf 2.588(026)}  &  0.070 &  0.717(043)   &  2.262(2)  &  1.557(09)  &  0.943(23)  &  0.917(61)  \\
0.959 &  2.564(143)  &  0.041 &  0.484(199)   &  2.312(2)  &  1.479(09)  &  1.149(28)  & -0.000(35)  \\
0.890 &  2.615(327)  &  0.110 &  0.804(172)   &  2.169(2)  &  1.425(07)  &  1.248(26)  &  0.217(37)  \\
0.971 &  3.241(832)  &  0.029 & 0.546(2.016)  &  2.904(4)  &  1.580(18)  &  0.723(42)  &  0.469(71)  \\
0.953 & 3.562(1.264) &  0.047 & 0.713(3.222)  &  2.994(4)  &  1.406(18)  &  0.888(39)  &  0.406(65)  \\
0.500 &  2.652(215)  &  0.500 &  2.652(538)   &  2.669(3)  &  1.653(15)  & 2.405(162)  & -0.000(158) \\ 
0.595 & 4.803(1.270) &  0.405 & 1.413(1.614)  &  2.269(2)  &  1.459(08)  &  1.414(35)  & -0.000(43)  \\
0.925 &  2.593(027)  &  0.075 &  0.750(041)   &  2.257(2)  &  1.560(09)  &  0.995(25)  &  0.729(55)  \\
0.812 &  1.048(014)  &  0.188 &  3.271(694)   &  1.232(1)  &  1.068(02)  &  1.079(06)  &  0.941(14)  \\
0.683 &  0.996(016)  &  0.317 &  2.039(123)   &  1.226(1)  &  1.089(02)  &  1.030(05)  &  1.070(15)  \\\hline
\end{tabular}
\caption[Best fit parameters and effective masses for Polyakov loops.]{\label{tab:poly}Best fit parameters and effective masses in lattice units for Polyakov loops after application of the variational procedure.}
\end{table}

\begin{table}
\centering
\begin{tabular}{|cccc|cccc|} \hline
\multicolumn{8}{|c|}{$0^+$ Box Operators} \\\hline
\multicolumn{4}{|c|}{Best Fit Parameters} & \multicolumn{4}{c|}{Effective Masses} \\\hline
 Amp1  &    Mass1  &  Amp2  &      Mass2  &   EffM1  &   EffM2  &    EffM3   &    EffM4    \\\hline
0.961  &  {\bf 1.216(052)}  &  0.039  &  5.415(721)    &  1.257(1)  &  1.201(02)  &   1.306(009)  &   1.502(039)   \\
0.793  &       1.702(189)  &  0.207  &  2.991(1.844)  &  1.864(1)  &  1.752(08)  &   1.716(042)  &  -0.000(042)   \\
0.500  &  {\bf 2.100(068)}  &  0.500  &  2.100(709)    &  2.103(2)  &  1.997(12)  &   0.000(297)  &  0.000(1.554)  \\
0.500  &       2.279(074)  &  0.500  &  2.279(074)    &  2.283(2)  &  1.998(15)  &   3.319(429)  &  -0.000(416)   \\
0.500  &  {\bf 2.259(070)}  &  0.500  &  2.259(070)    &  2.579(3)  &  2.476(34)  &   0.000(078)  &   0.000(137)   \\
0.551  &       4.464(738)  &  0.449  &  2.065(321)    &  2.761(4)  &  2.136(26)  & 5.951(1.983)  & -0.000(2.334)  \\
0.500  &       2.634(107)  &  0.500  &  2.634(107)    &  2.638(3)  &  2.175(26)  &   0.000(276)  &  -0.000(282)   \\
0.869  &       3.469(348)  &  0.131  &  1.486(1.166)  &  2.871(4)  &  2.017(27)  &   1.588(130)  &   0.634(278)   \\
0.500  &  {\bf 2.741(022)}  &  0.500  &  2.741(022)    &  2.740(3)  &  2.796(50)  &   0.000(096)  &  -0.000(087)   \\
0.996  &       2.732(164)  &  0.004  &  0.023(1.263)  &  2.669(3)  &  2.078(24)  &   0.626(045)  &   0.333(069)   \\
0.500  &       2.586(284)  &  0.500  &  2.586(594)    &  2.591(2)  &  2.098(23)  &   2.670(323)  &  -0.000(368)   \\
0.500  &       2.164(108)  &  0.500  &  2.164(1.127)  &  2.170(2)  &  1.939(12)  &   0.000(307)  &  0.000(1.421)  \\
0.500  &  {\bf 1.945(041)}  &  0.500  &  1.945(213)    &  2.112(2)  &  1.911(11)  &   2.306(112)  &  -0.000(108)   \\
0.913  &       1.241(023)  &  0.087  &  8.344(2.595)  &  1.333(1)  &  1.228(03)  &   1.304(010)  &   1.486(041)   \\\hline
\end{tabular}
\caption[Best fit parameters and effective masses for $0^+$ box operators.]{\label{tab:box}Best fit parameters and effective masses for $0^+$ box operators after application of the variational procedure.}
\end{table}

\begin{table}
\centering
\begin{tabular}{|cc|cc|} \hline
\multicolumn{4}{|c|}{Bar Operators} \\\hline
\multicolumn{2}{|c|}{$0^+$} & \multicolumn{2}{c|}{$2^+$} \\\hline
 Amp  &    Mass  &  Amp  &      Mass \\
0.518 &   {\bf 1.086(059)}  &  0.500  &  {\bf 1.932(044)} \\
0.500 &   {\bf 1.906(049)}  &  0.500  &  2.152(171) \\
0.996 &   {\bf 2.169(320)}  &  0.501  &  {\bf 2.145(212)} \\
0.628 &        1.341(055)  &  0.535  &  1.999(228) \\
0.599 &       6.174(1.974) &  0.500  &  {\bf 2.447(077)} \\
0.737 &        1.260(011)  &  0.903  &  2.969(185) \\\hline
\end{tabular}
\caption[Best fit parameters for $0^+$ and $2^+$ bar operators.]{\label{tab:bar02}Best fit parameters for $0^+$ and $2^+$ bar operators after application of the variational procedure.}
\end{table}

\begin{table}
\centering
\begin{tabular}{|cc|cc|} \hline
\multicolumn{4}{|c|}{$2^+$ 8-Bar Operators} \\\hline
\multicolumn{2}{|c|}{Dir1} & \multicolumn{2}{c|}{Dir2} \\\hline
 Amp  &    Mass  &  Amp  &      Mass \\
0.500  &  {\bf 2.045(029)} &  0.998  &  {\bf 2.038(122)} \\
0.926  &       2.553(388) &  0.936  &  2.576(514) \\
0.595  &       2.236(314) &  0.500  &  {\bf 2.581(132)} \\
0.528  &       2.151(293) &  0.500  &  2.960(059) \\
0.500  &       3.705(066) &  0.500  &  3.252(097) \\
0.847  &       2.896(275) &  0.500  &  2.781(086) \\
0.500  &  {\bf 2.527(104)} &  0.500  &  2.627(266) \\
0.500  &  {\bf 2.283(147)} &  0.500  &  {\bf 2.243(218)} \\\hline
\end{tabular}
\caption[Best fit parameters for $2^+$ 8-bar operators.]{\label{tab:8bar2}Best fit parameters for $2^+$ 8-bar operators after application of the variational procedure.}
\end{table}

\begin{table}
\centering
\begin{tabular}{|cc|cc|cc|} \hline
\multicolumn{6}{|c|}{8-Fold Operators} \\\hline
\multicolumn{2}{|c|}{$0^+$} & \multicolumn{2}{c|}{$2^+$} & \multicolumn{2}{c|}{$4^+$} \\\hline
 Amp  &    Mass  &  Amp  &      Mass  &  Amp  &      Mass \\
0.653  &  {\bf 1.148(045)}  &  0.500  &  {\bf 2.030(019)}  &  0.845  &  0.961(155) \\
0.500  &  {\bf 1.820(172)}  &  0.967  &  2.491(410)        &  0.500  &  2.045(037) \\
0.859  &  {\bf 2.128(038)}  &  0.500  &  {\bf 2.574(083)}  &  0.500  &  2.373(218) \\
0.500  &  2.041(125)        &  0.500  &  2.801(122)        &  0.500  &  {\bf 2.586(091)} \\
0.795  &  1.969(135)        &  0.726  &  4.197(716)        &  0.500  &  2.445(074) \\
0.859  &  {\bf 2.513(088)}  &  0.999  &  2.842(273)        &  0.791  &  4.360(761) \\
0.854  &  1.282(084)        &  0.500  &  2.164(096)        &  0.500  &  2.345(077) \\
0.759  &  1.284(050)        &  0.992  &  {\bf 2.364(320)}  &  0.773  &  2.107(063) \\\hline
\end{tabular}
\caption[Best fit parameters for $0^+$, $2^+$, and $4^+$ 8-fold operators.]{\label{tab:8fold024}Best fit parameters for $0^+$, $2^+$, and $4^+$ 8-fold operators after application of the variational procedure.}
\end{table}

\begin{table}
\centering
\begin{tabular}{|c|rrrrrrrr|} \hline
\multicolumn{9}{|c|}{8-Fold Overlaps} \\\hline
t=0 & \multicolumn{8}{c|}{$4^+$} \\\hline
\multirow{8}{*}{$0^+$}  
 &  0.891 &  0.118 &  0.238 & -0.144 & -0.299 &  0.158 & -0.100 &  0.719 \\
 &  0.377 & -0.640 & -0.288 &  0.344 &  0.460 & -0.137 &  0.442 & -0.309 \\
 &  0.503 &  0.326 &  0.495 & -0.313 & -0.542 &  0.229 & -0.228 &  0.384 \\
 &  0.406 &  0.389 &  0.506 & -0.601 & -0.847 &  0.170 & -0.264 &  0.344 \\
 &  0.400 & -0.455 & -0.538 &  0.610 &  0.625 & -0.150 &  0.267 & -0.299 \\
 &  0.467 & -0.348 & -0.421 &  0.282 &  0.504 & -0.172 &  0.266 & -0.366 \\
 &  0.860 & -0.416 & -0.361 &  0.326 &  0.495 & -0.195 &  0.301 & -0.706 \\
 &  0.559 & -0.192 &  0.068 &  0.033 & -0.049 &  0.050 &  0.109 &  0.455 \\\hline
t=1 & \multicolumn{8}{c|}{$4^+$} \\\hline
\multirow{8}{*}{$0^+$}  
 &  0.235 &  0.021 &  0.051 & -0.025 & -0.063 &  0.040 & -0.021 &  0.192 \\
 & -0.067 & -0.102 & -0.035 &  0.048 &  0.062 & -0.017 &  0.074 & -0.055 \\
 &  0.129 &  0.034 &  0.065 & -0.036 & -0.070 &  0.035 & -0.025 &  0.106 \\
 &  0.085 &  0.054 &  0.056 & -0.055 & -0.080 &  0.025 & -0.039 &  0.071 \\
 & -0.088 & -0.064 & -0.058 &  0.055 &  0.073 & -0.026 &  0.046 & -0.071 \\
 & -0.117 & -0.041 & -0.057 &  0.036 &  0.066 & -0.031 &  0.031 & -0.096 \\
 & -0.208 & -0.067 & -0.059 &  0.045 &  0.082 & -0.039 &  0.053 & -0.170 \\
 &  0.168 & -0.029 &  0.029 & -0.000 & -0.027 &  0.027 &  0.016 &  0.139 \\\hline
\end{tabular}
\caption[Timeslice overlaps for the $4^+$ 8-fold operators.]{\label{tab:8foldovrlaps}Timeslice overlaps for the $4^+$ 8-fold operators.  Rows correspond to the $0^+$ operators, and columns correspond to the $4^+$ operators.}
\end{table}

\begin{table}
\centering
\begin{tabular}{|cc|cc|cc|cc|} \hline
\multicolumn{8}{|c|}{12-Fold Operators} \\\hline
\multicolumn{2}{|c|}{$0^+$} & \multicolumn{2}{c|}{$2^+$} & \multicolumn{2}{c|}{$4^+$} & \multicolumn{2}{c|}{$6^+$} \\\hline
 Amp  &    Mass  &  Amp  &      Mass  &  Amp  &      Mass  &  Amp  &      Mass  \\
0.518 &  {\bf 1.108(049)}  & 0.500 &  {\bf 2.052(028)} &  0.500 &  1.483(074) & 0.500 &  2.122(043)  \\
0.871 &  {\bf 1.365(031)}  & 0.500 &  {\bf 2.261(038)} &  0.500 &  2.528(098) & 0.500 &  2.557(025)  \\
0.500 &  {\bf 2.159(084)}  & 0.770 &  2.404(211)       &  0.500 &  {\bf 2.645(087)} & 0.500 &  {\bf 3.021(110)}  \\
0.528 &  1.854(250)        & 0.500 &  {\bf 2.715(053)} &  0.500 &  2.658(112) & 0.570 &  2.510(276)  \\
0.585 &  3.440(566)        & 1.000 &  3.058(448)       &  0.500 &  3.020(046) & 0.999 &  3.108(872)  \\
0.558 &  1.403(041)        & 0.500 &  3.479(156)       &  0.710 &  5.169(848) & 0.500 &  3.522(374)  \\
0.564 & 6.204(1.214)       & 0.627 &  2.228(249)       &  0.500 &  3.284(070) & 0.500 & 3.779(1.038) \\ 
0.583 &  1.424(035)        & 0.739 &  2.064(352)       &  0.500 &  2.984(136) & 0.500 &  2.902(043)  \\
0.905 &  1.210(018)        & 0.925 &  2.952(582)       &  0.500 &  3.140(173) & 0.500 &  3.331(108)  \\
1.000 &  {\bf 1.864(100)}  & 0.936 &  2.329(313)       &  0.500 &  2.178(065) & 0.500 &  2.213(107)  \\\hline
\end{tabular}
\caption[Best fit parameters for $0^+$, $2^+$, $4^+$, and $6^+$ 12-fold operators.]{\label{tab:12fold0246}Best fit parameters for $0^+$, $2^+$, $4^+$, and $6^+$ 12-fold operators after application of the variational procedure.}
\end{table}

\begin{table}
\centering
\begin{tabular}{|c|rrrrrrrrrr|} \hline
\multicolumn{11}{|c|}{12-Fold Overlaps} \\\hline
t=0 & \multicolumn{10}{c|}{$4^+$} \\\hline
\multirow{10}{*}{$0^+$}  
 &  0.804 & -0.063 &  0.000 &  0.084 &  0.001 &  0.013 & -0.034 &  0.053 &  0.025 &  0.520 \\
 & -0.481 &  0.406 &  0.140 & -0.043 &  0.097 & -0.106 &  0.113 &  0.017 & -0.019 & -0.338 \\
 &  0.338 &  0.517 &  0.332 &  0.187 &  0.306 & -0.219 &  0.158 &  0.140 &  0.125 &  0.210 \\
 & -0.308 &  0.312 &  0.217 &  0.148 &  0.375 & -0.312 &  0.128 &  0.068 &  0.107 & -0.239 \\
 & -0.279 &  0.249 &  0.007 &  0.010 &  0.013 & -0.176 &  0.013 &  0.061 & -0.057 & -0.260 \\
 & -0.435 & -0.227 & -0.214 & -0.209 & -0.347 &  0.241 & -0.105 & -0.105 & -0.117 & -0.258 \\
 &  0.484 & -0.075 & -0.054 & -0.043 & -0.208 &  0.134 & -0.123 &  0.005 & -0.003 &  0.403 \\
 &  0.602 &  0.298 &  0.258 &  0.153 &  0.242 & -0.135 &  0.107 &  0.082 &  0.134 &  0.426 \\
 &  0.747 & -0.204 & -0.054 &  0.056 & -0.046 &  0.051 & -0.069 &  0.014 &  0.026 &  0.503 \\
 & -0.047 & -0.523 & -0.221 & -0.031 & -0.164 &  0.147 & -0.137 & -0.073 & -0.016 & -0.004 \\\hline
t=1 & \multicolumn{10}{c|}{$4^+$} \\\hline
\multirow{10}{*}{$0^+$}  
 &  0.211 & -0.011 & -0.002 &  0.026 & -0.003 &  0.004 & -0.006 &  0.017 &  0.004 &  0.135 \\
 & -0.143 &  0.057 &  0.016 & -0.018 &  0.012 & -0.015 &  0.015 & -0.004 & -0.008 & -0.097 \\
 &  0.068 &  0.050 &  0.028 &  0.021 &  0.023 & -0.021 &  0.013 &  0.017 &  0.009 &  0.043 \\
 & -0.071 &  0.020 &  0.012 &  0.006 &  0.022 & -0.019 &  0.008 &  0.003 &  0.006 & -0.047 \\
 & -0.074 &  0.043 &  0.004 & -0.006 &  0.005 & -0.012 &  0.006 &  0.004 & -0.009 & -0.053 \\
 & -0.123 & -0.004 & -0.010 & -0.027 & -0.016 &  0.012 & -0.001 & -0.016 & -0.010 & -0.080 \\
 &  0.101 &  0.010 &  0.005 &  0.008 & -0.005 &  0.004 & -0.000 &  0.008 & -0.000 &  0.065 \\
 &  0.138 &  0.024 &  0.019 &  0.024 &  0.015 & -0.011 &  0.006 &  0.017 &  0.010 &  0.090 \\
 &  0.199 & -0.026 & -0.007 &  0.023 & -0.006 &  0.008 & -0.009 &  0.013 &  0.005 &  0.129 \\
 &  0.009 & -0.071 & -0.022 &  0.001 & -0.016 &  0.019 & -0.016 & -0.010 &  0.006 &  0.012 \\\hline
\end{tabular}
\caption[Timeslice overlaps for the $4^+$ 12-fold operators.]{\label{tab:12fold04}Timeslice overlaps for the $4^+$ 12-fold operators.  Rows correspond to the $0^+$ operators, and columns correspond to the $4^+$ operators.}
\end{table}

\begin{table}
\centering
\begin{tabular}{|c|rrrrrrrrrr|} \hline
\multicolumn{11}{|c|}{12-Fold Overlaps} \\\hline
t=0 & \multicolumn{10}{c|}{$6^+$} \\\hline
\multirow{10}{*}{$2^+$}  
 &  0.651 & -0.078 &  0.002 & -0.095 & -0.158 &  0.156 &  0.074 &  0.204 &  0.203 &  0.476 \\
 & -0.650 &  0.463 & -0.018 &  0.207 &  0.179 & -0.220 & -0.125 & -0.426 & -0.224 & -0.390 \\
 &  0.208 & -0.565 &  0.116 & -0.383 & -0.355 & -0.004 &  0.119 &  0.407 &  0.116 &  0.110 \\
 & -0.383 &  0.704 & -0.122 &  0.444 &  0.141 & -0.205 & -0.065 & -0.629 & -0.146 & -0.172 \\
 &  0.147 & -0.283 &  0.074 & -0.371 & -0.488 & -0.237 &  0.084 &  0.214 &  0.132 &  0.127 \\
 &  0.028 &  0.009 & -0.062 &  0.068 &  0.179 &  0.439 & -0.067 &  0.029 &  0.002 &  0.013 \\
 &  0.554 & -0.400 &  0.015 & -0.182 & -0.020 &  0.248 &  0.066 &  0.462 &  0.196 &  0.355 \\
 & -0.620 &  0.122 &  0.061 &  0.053 &  0.139 & -0.137 & -0.112 & -0.174 & -0.257 & -0.457 \\
 & -0.108 &  0.488 & -0.117 &  0.323 &  0.306 &  0.078 & -0.111 & -0.320 & -0.052 & -0.041 \\
 &  0.561 & -0.070 &  0.003 & -0.132 & -0.228 &  0.082 &  0.087 &  0.173 &  0.211 &  0.437 \\\hline
t=1 & \multicolumn{10}{c|}{$6^+$} \\\hline
\multirow{10}{*}{$2^+$}  
 &  0.076 & -0.009 &  0.002 & -0.010 & -0.020 &  0.020 &  0.009 &  0.024 &  0.024 &  0.054 \\
 & -0.071 &  0.041 & -0.002 &  0.020 &  0.024 & -0.023 & -0.012 & -0.041 & -0.024 & -0.044 \\
 &  0.013 & -0.043 &  0.008 & -0.025 & -0.021 &  0.003 &  0.009 &  0.030 &  0.007 &  0.006 \\
 & -0.035 &  0.050 & -0.008 &  0.031 &  0.020 & -0.016 & -0.011 & -0.042 & -0.014 & -0.020 \\
 &  0.009 & -0.017 &  0.006 & -0.018 & -0.023 & -0.006 &  0.006 &  0.014 &  0.006 &  0.008 \\
 &  0.006 &  0.002 & -0.004 &  0.002 &  0.004 &  0.011 &  0.000 & -0.000 &  0.003 &  0.003 \\
 &  0.060 & -0.034 &  0.003 & -0.018 & -0.016 &  0.022 &  0.010 &  0.037 &  0.020 &  0.039 \\
 & -0.070 &  0.017 &  0.002 &  0.009 &  0.018 & -0.019 & -0.009 & -0.025 & -0.024 & -0.048 \\
 & -0.003 &  0.037 & -0.007 &  0.019 &  0.016 &  0.000 & -0.007 & -0.024 & -0.002 &  0.001 \\
 &  0.064 & -0.007 &  0.001 & -0.011 & -0.020 &  0.015 &  0.008 &  0.020 &  0.021 &  0.047 \\\hline
\end{tabular}
\caption[Timeslice overlaps for the $6^+$ 12-fold operators.]{\label{tab:12fold26}Timeslice overlaps for the $6^+$ 12-fold operators.  Rows correspond to the $2^+$ operators, and columns correspond to the $6^+$ operators.}
\end{table}

\clearpage

\end{document}